\documentclass[10 pt]{article}

\usepackage{amsmath,amstext,amsgen,amsbsy,amsopn,amsfonts,graphicx,amssymb,
rotating}

\begin{document}

\title{The Radially Vibrating Spherical Quantum Billiard}

\author{Mason A. Porter and Richard L. Liboff \\ \\ Center for Applied 
Mathematics \\ and \\ Schools of Electrical Engineering and Applied Physics \\ 
\\ Cornell University}

\date{July, 2000}

\maketitle

\begin{centering}
\section*{Abstract}
\end{centering}

\vspace{.1 in}

	We consider the radially vibrating spherical quantum billiard as a 
representative example of vibrating quantum billiards.  We derive necessary 
conditions for quantum chaos in $d$-term superposition states.  These 
conditions are symmetry relations corresponding to the relative quantum 
numbers of eigenstates considered pairwise.  In this discussion, we give 
special attention to eigenstates with null angular momentum (for which the 
aforementioned conditions are automatically satisfied).  When these necessary 
conditions are met, we observe numerically that there always exist parameter 
values for which the billiard behaves chaotically.  We focus our numerical 
studies on the ground and first excited states of the radially vibrating 
spherical quantum billiard with null angular momentum eigenstates.  We observe
 chaotic behavior in this configuration and thereby dispel the common belief 
that one must pass to the semiclassical ($\hbar \longrightarrow 0$) or high 
quantum number limits in order to meaningfully discuss quantum chaos.  The 
results in the present paper are also of practical import, as the radially 
vibrating spherical quantum billiard may be used as a model for the quantum 
dot nanostructure, the Fermi accelerating sphere, and intra-nuclear particle 
behavior.

\vspace{.1 in}

\section{Introduction}

	There has been considerable research in the last twenty years that 
seeks to marry quantum mechanics and dynamical systems theory into a coherent 
whole.\cite{casati,gutz}  In particular, the concept of quantum chaos extends 
the notions of classical Hamiltonian chaos to the quantum regime.  There are 
three types of quantum chaotic behavior:  ``quantized chaos'' (``quantum 
chaology''), ``semiquantum chaos,'' and genuine ``quantum chaos.''  Quantum 
chaology concerns the quantum structure of classically chaotic systems,  
semiquantum chaos refers to the chaotic dynamics of coupled classical and 
quantum systems, and genuine quantum chaos refers to chaotic dynamics of fully
 quantum systems.\cite{atomic}  No example of the third type of quantum chaos 
has been established, so the existence of such systems is an open question.

	In the present paper, we discuss semiquantum chaos in the context of 
vibrating billiard systems.  We review the recent results of Liboff and 
Porter\cite{sazim} and discuss them in further detail.  We treat a two-term 
Gal\"erkin projection (superposition state) of the radially vibrating sphere 
and prove that only 2-term superpositions whose normal modes have common 
rotational symmetry behave chaotically.  We extend this theorem to arbitrary 
superposition states by applying it pairwise.  In the proof of this theorem, 
we establish integrable (non-chaotic) behavior by showing that the evolution 
equations reduce to a two-dimensional autonomous dynamical system, whose 
non-chaotic properties are known.\cite{wiggins}  We then discuss two examples:
 one integrable superposition and one chaotic one.  We compute Poincar\'e maps
 and thereby reveal chaotic characteristics such as regions of ergodicity and 
KAM islands.

	We also discuss the present results with respect to the phenomenology 
of quantum chaos.  The chaotic behavior in the radius $a$ and conjugate 
momentum $P$ corresponds to classical Hamiltonian chaos.  The normal modes 
$\psi_{nlm}$ depend on the radius, so they exhibit quantum-mechanical wave 
chaos.  We also observe chaos in the Bloch variables $(x,y,z)$, which 
correspond to quantum-mechanical probabilities.  The dynamical equations 
describing the present system correspond to a two degree-of-freedom 
(\begin{itshape}dof\end{itshape}) Hamiltonian system, where one 
degree-of-freedom is classical (corresponding to the so-called one 
``degree-of-vibration'' (\begin{itshape}dov\end{itshape}) radial motion) and 
one is quantum-mechanical (corresponding to the coupling coefficient $\mu$).  
By coupling a single classical \begin{itshape}dof\end{itshape} (which is 
necessarily integrable) with a single quantum \begin{itshape}dof\end{itshape} 
(which must also be integrable), we obtain a genuinely chaotic system that 
provides an example of semiquantum chaos since it consists of a classical 
system coupled to a quantum one.  We remark that we do not need to pass to the
 semiclassical ($\hbar \longrightarrow 0$) or high quantum number limits in 
order to observe chaos, as is commonly considered requisite for a meaningful 
analysis of quantum chaos.\cite{gutz}  

	The radially vibrating spherical quantum billiard has several 
practical applications that complement its theoretical import. The most 
important one is that it may be used as a model for the quantum dot 
nanostructure.\cite{qdot}  At low temperatures, this microdevice component 
experiences vibrations due to zero-point motions, and at higher temperatures, 
it exhibits vibrations due to natural fluctuations.  Another application is 
that the radially vibrating spherical quantum billiard generalizes Fermi's 
``bouncing-ball model'' of cosmic ray acceleration.\cite{fermi}  Additionally,
 the radially vibrating spherical quantum billiard models the intradynamics of
 the nucleus, as the `liquid drop' and `collective' models of the nucleus 
include boundary vibrations.  Consequently, the importance of the radially 
vibrating spherical quantum billiard lies not only in its expansion of the 
theory of quantum chaos but also in its applicability to problems in nuclear, 
atomic, and mesoscopic physics.

\section{Statement of the Problem}

	The spherical quantum billiard addresses the quantum dynamics of a 
particle of mass $m_0$ confined to the interior of a spherical cavity of mass 
$M \gg m_0$ with smooth walls of radius $a$.  The radius vibrates in an a 
priori unspecified manner, so that $a \equiv a(t)$.  A two-component 
superposition state (Gal\"erkin projection) of this quantum billiard is given 
using Dirac notation by 
\begin{equation}
	|\psi(r,\theta,\phi,t;a(t))\rangle = A_1(t)|nlm,t\rangle + 
A_2(t)|n'l'm',t\rangle, \label{super}
\end{equation}
where $A_1(t)$ and $A_2(t)$ are complex amplitudes.  The numbers $\{n,l,m\}$ 
are, respectively, the principal, orbital, and azimuthal quantum numbers.  The
 eigenstates of the present system are products of spherical Bessel functions 
and spherical harmonics.\cite{liboff}  In coordinate representation,
\begin{equation}
	\langle \vec{r} | nlm,t\rangle = \psi_{nlm}(r,\theta,\phi,t;a(t)) = 
\sqrt{\frac{2}{a(t)^3}}\left(\frac{1}{j_{l + 1}(x_{ln})}\right)j_l 
\left(\frac{rx_{ln}}{a(t)}\right)Y_{lm}(\theta,\phi),
\end{equation}
where $x_{ln}$ is the $n$th zero of $j_l$, the spherical Bessel function of 
order $l$.  

	For the system at hand, the time-dependent Schr\"odinger equation is 
given by
\begin{equation}
	i\hbar \frac{\partial \psi}{\partial t} = K \psi, \hspace{.2 in} r 
\leq a(t), \label{sch}
\end{equation}
where the quantum-mechanical Hamiltonian $K$, the kinetic energy of the 
particle, is 
\begin{equation}
	K = -\frac{\hbar^2}{2m_0}\nabla^2.  
\end{equation}
The total Hamiltonian of the system is 
\begin{equation}
	H = \frac{P^2}{2M} + V + K, \label{totham}
\end{equation}
where $P$ is the momentum of the billiard boundary, and $V \equiv V(a) $ is 
the potential of the billiard surface.  The potential energy $V$ and kinetic 
energy $P^2/2M$ of the billiard walls are classical quantities, and the 
confined particle is quantum-mechanical.  For this semiquantum system, we 
utilize the Born-Oppenheimer approximation\cite{vibline}, so that only the 
particle kinetic energy $K$ is inserted into the Schr\"odinger equation 
(\ref{sch}).  In this adiabatic approximation, which is commonly used in 
mesoscopic physics, we are ignoring the effects of Berry phase.\cite{berry}

	Taking the expectation of (\ref{sch}) using the superposition state 
(\ref{super}) gives
\begin{gather}
	\left\langle  \psi \left|  -\frac{\hbar^2}{2m_0} 
\nabla^2\psi\right.\right\rangle = \frac{1}{a^2}\left[\epsilon_1 |A_1|^2 + 
\epsilon_2 |A_2|^2\right] \equiv K(A_1,A_2,a), \notag \\
	i\hbar \left\langle \psi \left| \frac{\partial \psi}{\partial t} 
\right.\right\rangle = i\hbar\left[\dot{A}_1A_1^* + \dot{A}_2A_2^* + 
\nu_{11}|A_1|^2 + \nu_{12}A_1A_2^* + \nu_{21}A_2A_1^* + 
\nu_{22}|A_2|^2\right], \label{expect}
\end{gather}
where the energies of the two terms are given by
\begin{gather}
	\epsilon_1 \equiv \frac{\hbar^2 x_{ln}^2}{2 m_0}, \hspace{.2 in} 
\epsilon_2 \equiv \frac{\hbar^2 x_{l'n'}^2}{2m_0}.
\end{gather}

\section{Integrable Configuration}

	Examining the superposition of $|100\rangle$ and $|110\rangle$ using 
(\ref{expect}) and orthogonality of spherical harmonics shows that $\nu_{11} =
 \nu_{12} = \nu_{21} = \nu_{22} = 0$.  (Note that $\nu_{11}$ and $\nu_{22}$ 
vanish no matter which eigenstates one considers.)  Equating the inner 
products (\ref{expect}) of both sides of the Schr\"odinger equation gives 
equations of motion for the complex amplitudes:
\begin{gather}
	i\dot{A}_1 = \frac{1}{\hbar a^2}\epsilon_1 A_1, \hspace{.2 in} 
i\dot{A}_2 = \frac{1}{\hbar a^2}\epsilon_2 A_2,
\end{gather}
which are integrated to yield
\begin{gather}
	A_1(t) = C_1 \exp\left[-\frac{i\epsilon_1}{\hbar}\int 
a^{-2}(t)dt\right], \hspace{.2 in} A_2(t) = C_2 
\exp\left[-\frac{i\epsilon_1}{\hbar}\int a^{-2}(t)dt\right]. \label{amp}
\end{gather}
From (\ref{amp}), one obtains a Hamiltonian in the radius $a$ and conjugate
 momentum $P$:
\begin{gather}
	H = \frac{P^2}{2M} + K(A_1,A_2,a) + V(a) = \frac{P^2}{2M} + \frac{1}
{a^2}\left[\epsilon_1 |C_1|^2 + \epsilon_2 |C_2|^2 \right] + V(a). 
\label{hamil}
\end{gather}
A Hamiltonian with no explicit time-dependence and one 
\begin{itshape}dof\end{itshape} corresponds to a two-dimensional autonomous 
system, which is known to be non-chaotic.\cite{gucken,wiggins}  When there are
 no coupling terms, the degree-of-freedom of the resulting Hamiltonian 
corresponds to the degree-of-vibration of the quantum billiard, which is a 
measure of the number of distance dimensions that undergo oscillations.  When 
a two-term superposition has a non-vanishing coupling coefficient, the number 
of degrees-of-freedom of the resulting Hamiltonian system is equal to the 
number of \begin{itshape}dov\end{itshape} of the billiard plus one.  In 
particular, this means that a superposition state of a quantum billiard with 
more than one \begin{itshape}dov\end{itshape} (such as the two 
\begin{itshape}dov\end{itshape} rectangular quantum billiard) is expected to 
behave chaotically even if every one of its coupling coefficients vanishes.

	Hamilton's equations for the present integrable configuration are 
\begin{gather}
	\dot{a} = \frac{P}{M} \equiv \frac{\partial H}{\partial P}, 
\hspace{.2 in} \dot{P} = -\frac{\partial V}{\partial a} + \frac{\lambda}{a^3} 
\equiv -\frac{\partial H}{\partial a}, \label{integ}
\end{gather}
where the energy parameter $\lambda$ is given by 
\begin{equation}
	\lambda \equiv 2\left( \epsilon_1 |C_1|^2 + \epsilon_1 |C_2|^2 
\right) > 0. \label{lam}
\end{equation}
The bifurcation structure of (\ref{integ}) has been studied for quartic 
potentials $V(a)$.\cite{bif} 

\section{Necessary Conditions for Chaos in k Coupled States}

	Consider the superposition
\begin{equation}
	\psi = A_1\psi_{q_1} + A_2\psi_{q_2} + \cdots + A_k\psi_{q_k}, 
\label{kstate}
\end{equation}
where $q_i \equiv (n_i,l_i,m_i)$ is a vector of quantum numbers.  If there
 does not exist a pair of normal modes in the $k$-state superposition 
(\ref{kstate}) with common angular momentum quantum numbers (i.e., there is no
 pair $\{i,i'\}$ such that $l_i = l_{i'}$ and $m_i = m_{i'}$), then inserting 
(\ref{kstate}) into the Schr\"odinger equation (\ref{sch}) returns a diagonal 
quadratic form
\begin{equation}
	\dot{A}_1A_1^* + \cdots + \dot{A}_kA_k^* = \nu_{11}|A_1|^2 + \cdots + 
\nu_{kk}|A_k|^2, \label{form2}
\end{equation}
as all the cross terms $\nu_{ij}A_iA_j^*$ have vanishing coupling coefficients
 $\nu_{ij}$ by orthogonality of spherical harmonics with different angular 
momenta.  The diagonal terms in (\ref{form2}) stem from the Laplacian.  As 
above, we obtain the Hamiltonian
\begin{equation}
	H(a,P) = \frac{P^2}{2M} + \frac{1}{a^2}\sum_{i = 1}^k \epsilon_i 
|C_i|^2 + V(a), \label{ha}
\end{equation}
where the $C_i$ are constants.  The superposition (\ref{kstate}) is 
non-chaotic, because the Hamiltonian (\ref{ha}) is autonomous with one 
\begin{itshape}dof\end{itshape}.

	We thus conclude that a necessary condition for chaotic behavior in 
an arbitrary finite superposition state of the radially vibrating spherical 
quantum billiard is that at least one pair of normal modes in the 
eigenfunction expansion have common angular momentum quantum numbers.  In 
particular, by considering small $n_i$ and $n_{i'}$, we obtain a chaotic 
superposition for eigenstates with small energies.  This even holds for some 
superpositions that include the ground state!  In most studies of quantum 
chaos, one must take the semiclassical ($\hbar \longrightarrow 0$) or high 
quantum-number limits in order to meaningfully study quantum 
chaos.\cite{atomic,gutz}  In such studies, termed ``quantum chaology,'' one 
considers the quantum signatures of classically chaotic systems in these 
regimes.  In the present system, on the other hand, we obtain genuinely 
chaotic behavior in a semiquantum system.  This phenomenon is often called 
``semiquantum chaos.''\cite{atomic}  

\section{Chaotic Configuration}

	As an example of a chaotic configuration of the radially vibrating 
spherical quantum billiard, consider the superposition state consisting of the
 ground and first excited states with null angular momentum
\begin{equation}
	|\psi(n,l,m)\rangle = A_1|100\rangle + A_2|200\rangle, \label{supnull}
\end{equation}
which gives the wavefunction
\begin{equation}
	\psi(r,t) = A_1(t)\alpha_1\psi_1(r,t)e^{-i \frac{E_1 t}{\hbar}} + 
A_2(t)\alpha_2\psi_2(r,t)e^{-i \frac{E_2 t}{\hbar}}, 
\end{equation}
where 
\begin{equation}
	\psi_n(r,t) = j_0\left(\frac{n\pi r}{a(t)}\right), \hspace{.2 in} 
j_0(x) = \frac{sin(x)}{x}, \hspace{.2 in} \alpha_n = 
\frac{\sqrt{2}}{a^\frac{3}{2}j_1(n\pi)}.
\end{equation}
The superposition (\ref{supnull}) has a coupling coefficient $\mu \equiv 
\mu_{12} = 4/3$.  

	Equating coefficients in the quadratic form (\ref{expect}) gives the 
matrix equation
\begin{equation}
	i\dot{A}_n = \sum_{k = 1}^2 D_{nk}A_k, \label{cplx}
\end{equation}
where $D \equiv (D_{ij})$ is the Hermitian matrix
\begin{equation}
	D =  
	\begin{pmatrix}
		\frac{\epsilon_1}{\hbar a^2} & -i \mu \frac{\dot{a}}{a} \\
		i \mu \frac{\dot{a}}{a} & \frac{\epsilon_2}{\hbar a^2}
	\end{pmatrix}
   	\text{ ,} \label{dd}
\end{equation}	
and the energy coefficient $\epsilon_j$ is given by
\begin{equation}
	\epsilon_j \equiv \frac{ (j \pi \hbar)^2}{2 m_0}, \hspace{.2 in} j
 \in \{1,2\}.
\end{equation}
Defining the density matrix\cite{liboff} by $\rho_{qn} = A_q A_n^*$, 
introducing (dimensionless) Bloch variables $x \equiv \rho_{12} + \rho_{21}$, 
$y \equiv i(\rho_{21} - \rho_{12})$, and $z \equiv \rho_{22} - \rho_{11}$, and
 using (\ref{cplx}), we obtain the following three differential equations:
\begin{gather}
	\dot{x} = -\frac{\omega_0 y}{a^2} - \frac{2 \mu P z}{Ma}, 
\hspace{.2 in} \dot{y} = \frac{\omega_0 x}{a^2}, \hspace{.2 in} \dot{z} = 
\frac{2 \mu P x}{Ma}. \label{bev}
\end{gather}
In these equations, $\omega_0 \equiv (\epsilon_2 - \epsilon_1)/\hbar$.  
Rewriting the kinetic energy $K(A_1,A_2,a)$ in terms of the Bloch variable $z$
 gives
\begin{equation}
	K(z,a) = \frac{1}{a^2}(\epsilon_+ + z\epsilon_-), \hspace{.2 in}
\epsilon_\pm \equiv \frac{1}{2}(\epsilon_2 \pm \epsilon_1).
\end{equation}
Inserting $K(z,a)$ into the Hamiltonian (\ref{totham}) gives Hamilton's 
equations:
\begin{gather}
	\dot{a} = \frac{P}{M}, \hspace{.2 in} \dot{P} = 
-\frac{\partial V}{\partial a} + \frac{2}{a^3}\left[\epsilon_+ + 
\epsilon_-(z - \mu x)\right]. \label{class}
\end{gather}
Equations (\ref{bev}) and (\ref{class}) constitute a set of five coupled 
nonlinear ordinary differential equations, which can be shown to be equivalent
 to a two degree-of-freedom Hamiltonian system.  The constants of motion of 
the present system are the radius of the Bloch sphere
\begin{equation}
	x^2 + y^2 + z^2 \equiv |A_1|^2 + |A_2|^2 = 1
\end{equation}
and the energy (total Hamiltonian)
\begin{equation}
	H = \frac{P^2}{2M} + V(a) + K(z,a).
\end{equation}
The equilibria of equations (\ref{bev}, \ref{class}) satisfy $x = y = 0$, $z =
 \pm 1$, $P = 0$, and $a = a_\pm$, where $a_\pm$ satisfies the equation
\begin{equation}
	\frac{\partial V}{\partial a} = \frac{2}{a^3} (\epsilon_+ \pm 
\epsilon_-), \label{radius}
\end{equation}
where the subscript of $a_\pm$ corresponds to the sign of $z = \pm 1$.  
Assuming that $V(a) + K(z,a)$ has a single minimum in $a$, these equilibria 
are elliptic.\cite{sazim,bif}  (That is, every eigenvalue of the Jacobian of 
the linearized system is purely imaginary.)  For the harmonic potential
\begin{equation}
	V(a) = \frac{V_0}{a_0^2}(a - a_0)^2,
\end{equation}
the total energy of the billiard's boundary is given by
\begin{equation}
	V(a) + K(a) = \frac{V_0}{a_0^2}(a - a_0)^2 + \frac{\epsilon_+ + 
z\epsilon_-}{a^2}.
\end{equation}
With this choice of potential, equation (\ref{radius}) becomes
\begin{equation}
	a - a_0 = \frac{a_0^2\epsilon_\pm}{V_0a^3}. \label{spec}
\end{equation}
The solutions of (\ref{spec}) for the equilibrium radii $a_\pm$ correspond to 
the $\epsilon_\pm$ values.  One computes that $a_+ \geq a_- \geq a_0$, from 
which it follows that $a_- \leq a(0) \leq a_+$, so $a(t)$ remains bounded in 
the interval $[a_-,a_+]$.\cite{sazim}

	Now consider the superposition of the first $k$ null angular momentum 
eigenstates,
\begin{equation}
	\psi^{(k)}(r) = \sum_{n=1}^k A_n(t) \alpha_n \psi_n(r,t).
\end{equation}
In order to analyze this configuration, one first examines the 2-term 
superposition
\begin{equation}
	\psi_{nq} = A_n(t) \alpha_n \psi_n(r,t) + A_q(t) \alpha_q \psi_q(r,t),
 \hspace{.2 in} n < q,
\end{equation}
and then superposes the couplings one obtains from each $\psi_{nq}$ as $n$ and
 $q$ run from $1$ to $k$ in order to obtain dynamical equations for the 
amplitudes $A_i$.  One computes the coupling coefficients $\mu_{nq}$ to be
\begin{equation}
	\mu_{nq} = 2 \frac{q n}{a(t)(n + q)(q - n)}, \hspace{.2 in} n \neq q.
\end{equation}
The dynamical equations for $A_i$ are described by a $k \times k$ matrix and 
are a straightforward generalization of (\ref{cplx}, \ref{dd}).

\subsection{Numerical simulations}

	The analysis for 2-term superpositions of null angular momentum 
eigenstates follows that for the general case.\cite{sazim}  In the present 
case, the necessary conditions for chaotic behavior are satisfied 
automatically, because the quantum numbers $m$ and $l$ vanish for every normal
 mode under consideration.  Consequently, any $k$-term superposition 
($k \geq 2$) of null angular momentum eigenstates exhibits chaotic behavior.  
We consider numerical simulations for the coupling of the ground state and 
first excited state of a billiard residing in a harmonic potential.

\vspace{-.6 in}

\begin{figure}[htb] 
	\begin{centering}
		\leavevmode
		\includegraphics[width = 2 in, height = 2.4 in]{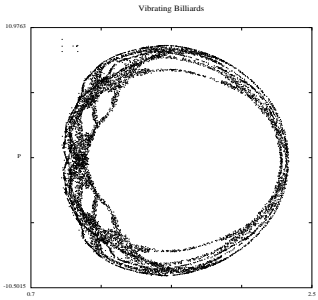}

\vspace{-.5 in}

		\caption{Poincar\'e Section $(x = 0)$ in the $(a,P)$-plane 
illustrating that not all invariant tori are destroyed in the present 
configruation.} \label{fig1}
	\end{centering}
\end{figure}

\begin{figure}[htb] 
	\begin{centering}
		\leavevmode
		\includegraphics[width = 2 in, height = 2.4 in]{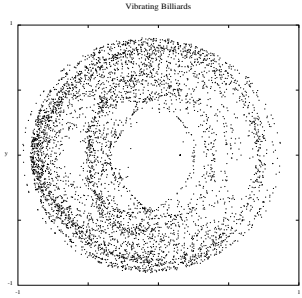}

\vspace{-.5 in}

		\caption{Poincar\'e Section $(P = 0)$ of the Bloch sphere 
projected onto the $(x,y)$-plane.  The structure in this diagram likewise 
illustrates the survival of some invariant tori.} \label{fig2}
	\end{centering}
\end{figure}

	Figure \ref{fig1} shows a Poincar\'e map in the $(a,P)$-plane 
corresponding to $x = 0$, and Figure \ref{fig2} shows a Poincar\'e section 
projected onto the $(x,y)$-plane for $P = 0$.  For each of these two plots, we 
used the parameter values $\hbar = 1$, $M = 10$, $m = 1$, $V_0/a_0^2 = 5$, and
 $a_0 = 1.25$.  The initial conditions for the two figures are $x(0) = 
\sin(0.95 \pi) \approx 0.156434$, $y(0) = 0$, $z(0) = \cos(0.95 \pi) \approx 
-0.987688$, $a(0) \approx 1.6$, and $P(0) \approx 9.45$.

\begin{figure}[htb] 
	\begin{centering}
		\leavevmode
		\includegraphics[width = 2 in, height = 2.4 in]{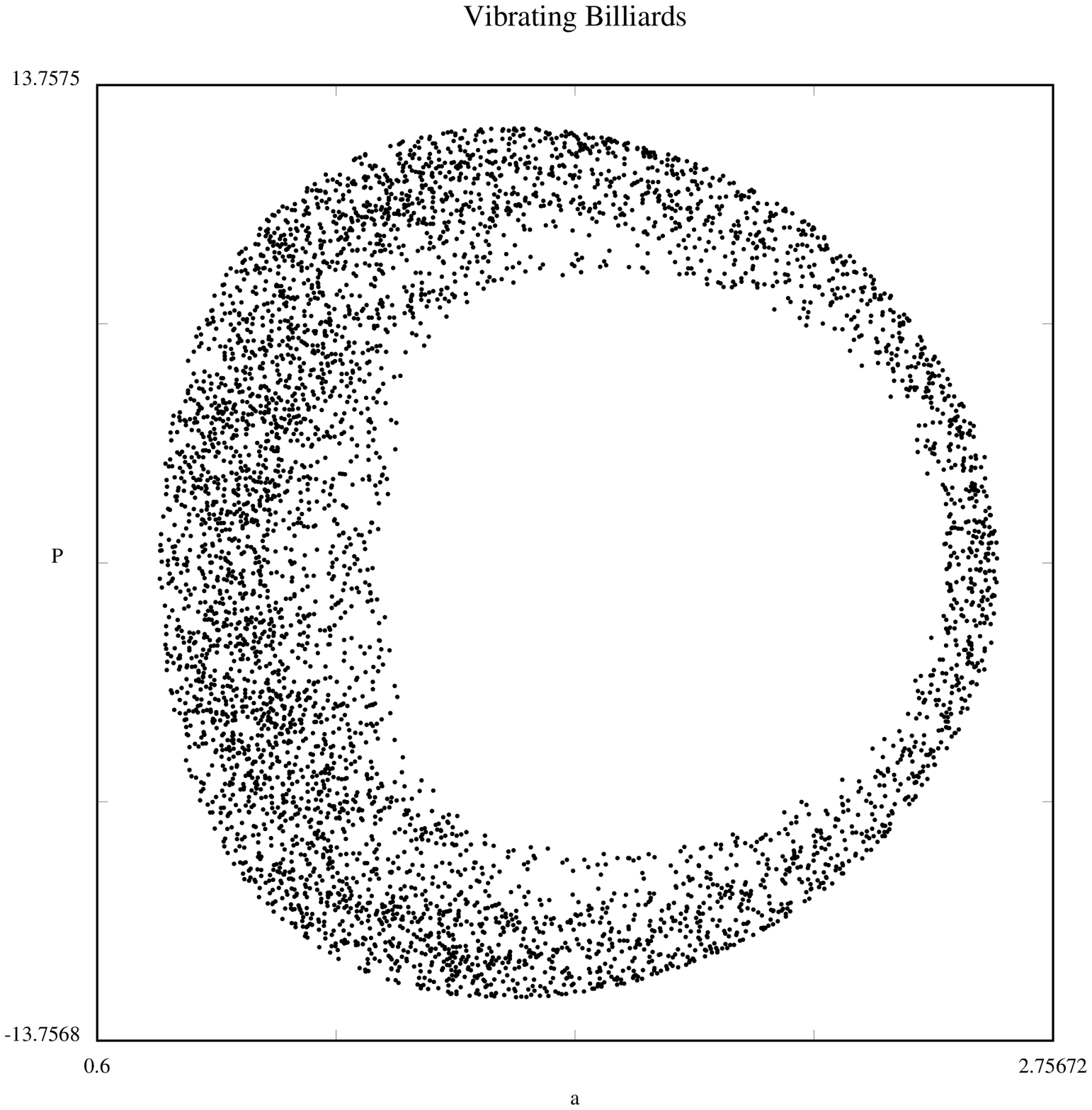}

\vspace{-.5 in}

		\caption{Poincar\'e Section $(x = 0)$ in the $(a,P)$-plane 
for slightly different initial conditions in which fewer invariant tori 
persist.  This is in accord with KAM theory.} \label{fig3}
	\end{centering}
\end{figure}

\begin{figure}[htb] 
	\begin{centering}
		\leavevmode
		\includegraphics[width = 2 in, height = 2.4 in]{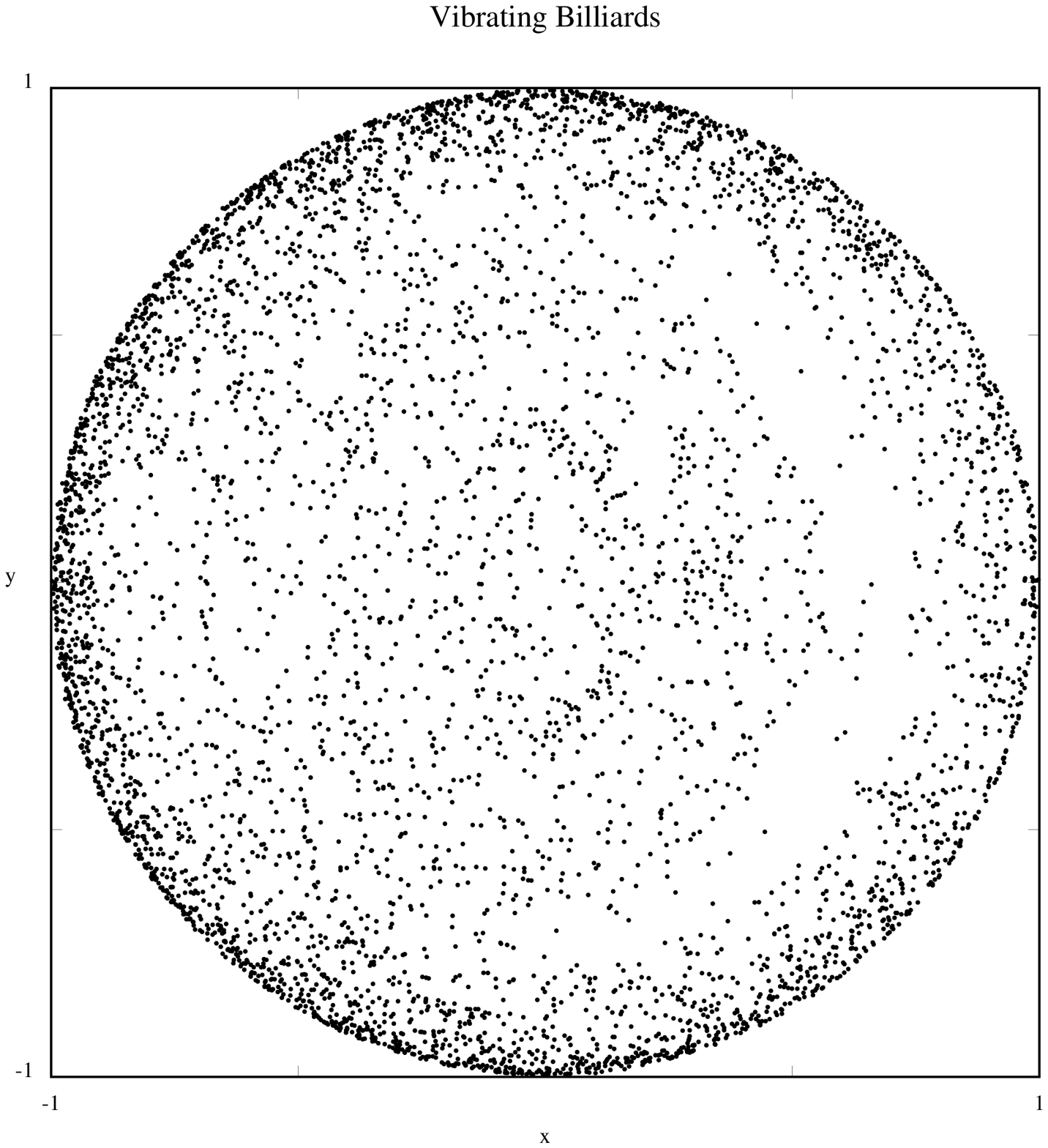}

\vspace{-.5 in}

		\caption{Poincar\'e Section $(P = 0)$ of the Bloch sphere 
projected onto the $(x,y)$-plane.  The initial conditions in this plot 
are the same as those in Figure \ref{fig3}.} \label{fig4}
	\end{centering}
\end{figure}

	The chaotic behavior of this configuration is evident in both plots, 
although there is clearly still some non-chaotic structure present.  In the 
language of KAM theory, some of the nonresonant tori persist for the present 
choice of initial conditions\cite{gucken,wiggins}.  One may also choose 
initial conditions corresponding to a different level of persistence of the 
resonant tori.  For example, Figure \ref{fig3} shows an $x = 0$ Poincar\'e map
 in the $(a,P)$-plane with the same initial conditions and parameter values as
 above, except $a(0) = 3$ and $P(0) = 10$.  Figure \ref{fig4} shows a $P = 0$ 
Poincar\'e map in the $(x,y)$-plane for these conditions.  There are fewer 
invariant tori in these two figures than there are in Figures 
\ref{fig1}--\ref{fig2}.  

\section{Phenomenology}

	In contrast to the present quantum-mechanical context, we note 
that for the classical radially vibrating spherical billiard, every orbit with
 null angular momentum is well-defined and invariant under radial oscillations
 of the boundary.\cite{corr}  Due to conservation of angular momentum, one 
finds that for the stationary spherical classical billiard, the enclosed 
particle sweeps out an annular domain of constant inner radius.\cite{sazim}  
Vibration of the wall of the sphere destroys this constant, and chaotic motion
 is expected to develop.  In the present quantum-mechanical context, we note 
that null angular momentum wavefunctions are composed only of spherical waves.
  The nodal surfaces of these wavefunctions are likewise spherical.  
Accordingly, the chaotic signature of this configuration in real space is the 
sequence of intersections with a fixed radius that nodal surfaces make at any 
instant subsequent to a number of transversal times.\cite{sazim}  This latter 
condition is consistent with the standard long-time behavior of chaotic 
dynamical systems.\cite{gutz}

	In the language of Bl\"umel and Reinhardt\cite{atomic}, vibrating 
quantum billiards are an example of semiquantum chaos.  One has a classical 
system (the walls of the billiard) coupled to a quantum-mechanical one (the
 enclosed particle).  Considered individually, each of these subsystems is 
integrable, as each contributes a single \begin{itshape}dof\end{itshape}.  
When they are coupled, however, one observes chaotic behavior in both of them.
  The classical variables $(a,P)$ exhibit Hamiltonian chaos, whereas the 
quantum subsystem $(x,y,z)$ is truly quantum chaotic.  Chaos on the Bloch 
sphere is an example of quantum chaos because the Bloch variables $(x,y,z)$ 
correspond to the quantum probabilities of the wavefunction.  Additionally, 
each individual normal mode $\psi_n$ depends on the radius $a(t)$, so each 
eigenfunction is an example of quantum-mechanical wave chaos for chaotic 
configurations of the billiard.  Moreover, because the evolution of the 
probabilities $|A_i|^2$ is chaotic, the wavefunction $\psi$ in the present 
configuration is a \begin{itshape}chaotic combination\end{itshape} of chaotic 
normal modes.  Finally, we note that if we quantize the motion of the billiard
 walls, we would obtain a higher-dimensional, fully-quantized system that 
exhibits so-called quantized chaos.\cite{atomic}  In particular, the fully 
quantized version of the present system would require passage to the 
semiclassical limit in order to observe quantum signatures of classical chaos.

\section{Conclusions}

	We considered the radially vibrating spherical quantum billiard as a 
representative example of vibrating quantum billiards.  We derived necessary 
conditions for quantum chaos in $k$-term superposition states.  We gave 
special attention to eigenstates with null angular momentum, for which these 
conditions are automatically satisfied.  We examined a numerical simulation of
 the superposition of the ground and first null angular momentum excited 
states.  We observed chaotic behavior in this configuration, thereby 
dispelling the common belief that one is required to pass to the semiclassical
 ($\hbar \longrightarrow 0$) or high quantum number limits in order to 
meaningfully study quantum chaos.

\section{Acknowledgements}

	We express our gratitude to the conference organizers (especially 
Joshua Du) for their excellent work.  We also thank the speakers for the 
wonderful talks they gave and the academic committee for their role in the
 organization process.

\vspace{.1 in}

\bibliographystyle{plain}
\bibliography{ref}

\end{document}